# Modelling correlated ordinal data by random-effects logistic regression models: simulation and application

Ali Reza Fotouhi
University of the Fraser Valley
33844 King Road, Abbotsford
BC, V2S 7M8,Canada
E-mail: ali.fotouhi@ufv.ca
Fax: +1 (604) 855-7614

Theresa Mulder
University of the Fraser Valley
33844 King Road, Abbotsford
BC, V2S 7M8,Canada
E-mail: theresa.mulder@ufv.ca

**Abstract:**

Ordered categorical data frequently arise in the analysis of biomedical, agricultural, and social sciences data. The logistic regression model is attractive in analyzing ordered categorical data because of its use in interpretation of a parameter estimate. The ordered responses may be clustered and the subjects within the clusters may be positively correlated. To accommodate this correlation we add a random component to the linear predictor of each clustered response. This article presents and compares random-effects logistic regression models for analyzing ordered categorical data. The proposed models are applied to an agricultural experimental data. In order to assess the performance of the random-effects and homogeneous models we perform a simulation study. Our analysis, application to real data and simulation, show that the probability of the individual categories are estimated poorly in homogeneous models. The random-effects models fit the data statistically significant and estimate the probability of the individual categories more precisely.

*Keywords:* Ordered categorical data; random-effects; logistic regression; simulation

## 1. Introduction

In many statistical studies the responses are recorded on an ordinal scale. An example is the strawberry data analyzed by Jansen [1], Crouchley [2], and Fotouhi and Dudzik [3]. Twelve populations of strawberry plants are compared through a randomized block design.  The 12 populations were obtained by crossing 3 male parents with 4 female parents. Each plant was assigned to one of three ordered categories, representing the level of damage caused by the fungus. In this example the ordered responses are clustered and the responses on the subjects within the same cluster may be positively correlated. A commonly used method to accommodate this correlation is to add a random component to the linear predictors of each clustered response. Moreover, the random-effect in a cluster may be variable across the ordered categories. In this case, a vector of random-effects can capture the heterogeneity within both the clusters and the categories. Random-effects component can also control any possibly omitted variables and overdispersion (Fotouhi [4] and [5]).

Sometimes categories are the result of grouping continuous data. However, this continuous random variable may not be observable. In such a case, ordered categorical data can be modeled by using a latent variable. A latent variable model provides a link between the ordinal scale of measurement and a linear scale on which treatments are supposed to act. If we assume that the latent variable has a normal distribution then the categorical variable has a probit link model. Assuming logistic distribution leads to logit link model, assuming an extreme value of minimums leads to a complementary log-log link model, and assuming an extreme value of maximums leads to a log-log link model. Jansen [1] has analyzed the strawberry data by using a probit link model. Crouchley [2] has analyzed the same data by using a complimentary log-log link. Fotouhi and Dudzik [3] have compared different link functions by using a latent variable random-effects model. However, in some cases the ordered categorical data may not be the result of grouping continuous data. The logistic regression has the proportional odds model, the adjacent categories logit model, and the continuation ratio logit model as its special cases. Dobson and Barnet [6] and Agresti [7] have discussed the details of these models for analyzing ordered categorical data.

In this article we introduce random-effects models for analyzing ordered categorical data when heterogeneity exists between both the clusters and the categories. We present a simulation study and an application to real data and compare the proportional odds, adjacent categories logit, and continuation ratio logit models. We also illustrate the importance of modeling the random-effects through simulation and application to the completely randomized block designed strawberry data.

In section 2 we introduce the models, in section 3 we apply the proposed models to the strawberry data (Jansen [1]), in section 4 we report some simulation results, and in section 5 we present some discussions.

## 2. Models

Following Dobson and Barnet [6] the three ordinal logistic regression models for $K$ ordered categories are
Proportional odds model:
$$\log \frac{\pi_{i1}+\pi_{i2}+\cdots+\pi_{ik}}{\pi_{i,k+1}+\pi_{i,k+2}+\cdots+\pi_{iK}} = \delta_{ik}, k = 1,2,\ldots,K-1 \qquad (1)$$
Adjacent categories logit model:
$$\log \frac{\pi_{ik}}{\pi_{i,k+1}} = \delta_{ik}, k = 1,2,\ldots,K-1 \qquad (2)$$
Continuation ratio logit model:
$$\log \frac{\pi_{ik}}{\pi_{i,k+1}+\pi_{i,k+2}+\cdots+\pi_{i,K}} = \delta_{ik}, k = 1,2,\ldots,K-1 \qquad (3)$$
where $\delta_{ik} = \beta_{0k} + \beta_1 x_{i1} + \cdots + \beta_{p-1} x_{i,p-1}, i = 1,2,\ldots,I$ is the linear predictor and $\pi_{i1}, \pi_{i2}, \ldots, \pi_{iK}$ with $\sum_{k=1}^{K} \pi_{ik} = 1$ represent the probability of the categories.

It is assumed that only the intercept depends on the category in these models. As different odds are used in equations 1, 2, and 3, the interpretation of the parameters are different. It is also assumed that the intercept $\beta_{0k}$ and the explanatory variables $x_i$ are independent. This assumption is desirable for many applications but it requires the strong assumption that wherever the cut points that separate categories are, the desired odds for a one unit change in $x_i$ is the same for all response categories while other explanatory variables are kept fixed.

To drive the likelihood function let $y_{ik}$ denote the number of outcomes in category $k$ and within cluster $i$. Conditional on the explanatory variables $x_i$ and the random effect $\varepsilon_i$, the likelihood of observing $y_{i1}, y_{i2}, \ldots, y_{iK}$ in cluster $i$ is given by

$$f(y_{i1}, y_{i2}, \ldots, y_{iK}|x_i) = \frac{N!}{y_{i1}!\, y_{i2}!, \ldots, y_{iK}!} \pi_{i1}^{y_{i1}} \pi_{i2}^{y_{i2}}, \ldots, \pi_{iK}^{y_{iK}} \qquad (4)$$

where $\pi_{ik}$ can be calculated from equations 1, 2, or 3 depending on the desired link function. In order to control for possible omitted variables and between cluster heterogeneity we add a random component $\varepsilon_i$ to the linear predictor $\delta_{ik}$. The conditional random-effects likelihood function for cluster $i$, conditional on $x_i$ and $\varepsilon_i$, is given by

$$f(y_{i1}, y_{i2}, \ldots, y_{iK}|x_i, \varepsilon_i) = \frac{N!}{y_{i1}!\, y_{i2}!, \ldots, y_{iK}!} \pi_{i1}^{y_{i1}} \pi_{i2}^{y_{i2}}, \ldots, \pi_{iK}^{y_{iK}} \qquad (5)$$

where $\delta_{ik} = \beta_{0k} + \beta_1 x_{i1} + \cdots + \beta_{p-1} x_{i,p-1} + \varepsilon_i$, $i = 1,2,\ldots,I$.

The normal distribution with mean 0 and standard deviation $\sigma$ is widely used as a parametric distribution for the random-effects. Integrating out the random-effects $\varepsilon_i$ gives the marginal likelihood function for the $i^{th}$ cluster as

$$L_i(\beta, \sigma) = \int_{-\infty}^{+\infty} f(y_{i1}, y_{i2}, \ldots, y_{iK}|x_i, \varepsilon_i) dG(\varepsilon_i). \qquad (6)$$

where $G$ is the cumulative distribution of the normal distribution with mean zero and standard deviation $\sigma$. The overall likelihood function over $N$ clusters is

$$L(\beta, \sigma) = \prod_{i=1}^{N} \int_{-\infty}^{+\infty} f(y_{i1}, y_{i2}, \ldots, y_{iK}|x_i, \varepsilon_i) dG(\varepsilon_i) \qquad (7)$$

and the log-likelihood function is given by

$$l(\beta, \sigma) = \sum_{i=1}^{N} \log \int_{-\infty}^{+\infty} f(y_{i1}, y_{i2}, \ldots, y_{iK}|x_i, \varepsilon_i) dG(\varepsilon_i). \qquad (8)$$

Three goodness of fit statistics; Pearson chi square ($\chi^2 = \sum_{i=1}^{N} \frac{(O_i - E_i)^2}{E_i}$), likelihood ratio chi square statistic ($C = 2[l(\beta) - l(\beta_{min})]$), and Akaike Information Criterion ($AIC = -2l(\hat{\pi}) + 2p$) are used to compare model fits. $\chi^2$ is used for testing if the model fit the data well, $C$ is used for comparison between the fitted model and the intercept model, and $AIC$ is used for comparison between models which are not nested. $\chi^2$ has a chi-squared distribution with $N - p$ degrees of freedom where $N$ is $(K - 1)$ times of the covariate pattern subtracted by the number of the parameters estimated by the model. $C$ has a chi-squared distribution with $p - (K - 1)$ degrees of freedom.

For the ordinal logistic model assuming normally distributed random-effects, the intraclass correlation (*ICC*) equals $\sigma^2/(\sigma^2 + \pi^2/3)$, where the latter term in the denominator represents the variance of the underlying latent response tendency. For the logistic model, this variable is assumed to be distributed as a standard logistic distribution with variance equal to $\pi^2/3$. This statistic measures the relative variability that is explained by the random-effects component. It can be shown that *ICC* is the correlation between two observations in a cluster in a latent variable framework.

## 3. Application

The grouped data from Jansen [1] are reported in Table 1 and are from an experiment concerning resistance against the fungus phytophtora fragariae in seedling populations of strawberries. Twelve populations of strawberries were tested in randomized blocks experiment with four blocks. Each block had similar soil and environmental characteristics. The plots consisted of 10 plants, but in some cases only 9 plants are observed. At the end of the experiment each plant was assigned to one of three ordered categories, representing the level of damage caused by the fungus. The 12 populations were obtained by crossing 3 male parents with 4 female parents. The 48 plots can be defined as the combinations of the 3 levels of male parents, 4 levels of female parents, and 4 blocks. We introduce two indicator variables for males (level one is the reference level), three indicator variables for females (level one is the reference level), and three indicator variables for blocks (block one is the reference block). The linear predictor, right hand side in equations 1, 2, and 3, is of the form

$$\delta_{ik} = c_k + m_2 x_{i1} + m_3 x_{i2} + f_2 x_{i3} + f_3 x_{i4} + f_4 x_{i5} + b_2 x_{i6} + b_3 x_{i7} + b_4 x_{i8}$$
$$+ \varepsilon_i, i = 1,2, \dots 48 \ ; k = 1,2. \quad (9)$$

Where $c, m, f, b$ present the intercept, male's effect, female's effect, and block's effect respectively.

Jansen [1] noted that variation between and within plots is partly genetic and partly environmental, as plants from the same cross are not genetically identical. He proposed the existence of an independent and identically normally distributed cluster-specific random-effects to account for the possibility that some important between plot differences have been left out of the linear predictors of the model. He has compared a homogeneous model with the normal random-effects model and concluded that the normal random-effects model has a better fit to the data. Crouchley [2] has analyzed the same data by using the complimentary log-log link. Fotouhi and Dudzik [3] have compared the probit, complementary log-log, logit, and log-log link functions by using a latent variable random-effects model.

We analyze this data by using three logistic models presented in equations 1, 2, and 3 with the linear predictor in equation 9 with and without random-effects. The parameters are estimated by maximizing the likelihood function presented in equation 8. We use procedure NLMIXED from SAS 9.3. The results are reported in Tables 2, 3, and 4.

The standard error of the random-effects is estimated significantly positive in all models. This suggests that the random-effects models have captured the between-plots heterogeneity and are therefore preferred to homogenous (no random-effects) models. The intraclass correlation ($ICC$) is estimated significantly positive in random-effects models that confirm the existence of positive correlation between observations within a cluster. The Akaike Information Criterion ($AIC$) suggests that the random-effects models fit the data better than models with no random effects (homogenous models). The Pearson chi-squared statistic $\chi^2$ with $df = 86$ degrees of freedom (calculated as
$df = (K - 1)(covariate\ pattern) - p$) and high $p - value$ shows that all random-effects models fit the strawberry data well, while none of the homogeneous models fit the data well. The $\chi^2$ statistic is estimated almost equal in three random-effects models, which indicates that these models estimate the probability of the individual categories almost equal. The likelihood ratio chi-squared statistic $C$ with $df = 8$ degrees of freedom (calculated as the difference between the number of the parameters in the fitted model and number of the parameter in the intercept model shows that all of the models fit the data significantly better than the minimal model (intercept model) at significance level of 5%.

Focusing on the fitted random-effects models, the directions of the estimates are consistent in all of the models, consistent with the estimates of Crouchley's [2] complementary log-log model, and with the estimates reported by Fotouhi and Dudzik [3] using probit, logit, complementary log-log, and log-log link functions. However, the magnitudes of the estimates in tables 2, 3, and 4 are rather different.

This makes sense as the structure of the models and the interpretation of the parameters are different among the models introduced in equations 1, 2, and 3. The male's effect is not significant by using any of these models, the fourth block's effect is not significantly different from the first block, and the female parent's effects are significant. The intercepts are estimated significantly negative in all of the models. The adjacent categories and continuation ratio logistic models are easier for interpretation than the proportional odds logistic model if the probabilities for the individual categories are of interest (Agresti [8]). As the continuation ratio logit random-effects model has produced the smallest $\chi^2$ statistic, it may be recommended for the interpretation of the probabilities for individual categories.

We extend the analysis by considering $\varepsilon_{ik}$ instead of $\varepsilon_i$ in equation 9. This is equivalent to assuming that the intercepts in the linear predictor presented in equation 9 are random with different distributions. We assume that $\varepsilon_{ik} = (\varepsilon_{i1}, \varepsilon_{i2})$ has a bivariate normal distribution with mean 0 and covariance matrix

$$V = \begin{bmatrix} \sigma_1^2 & \rho\sigma_1\sigma_2 \\ \rho\sigma_1\sigma_2 & \sigma_2^2 \end{bmatrix}.$$

We mentioned that the variation between and within plots is partly genetic and partly environmental. It is also possible that the probability of assigning a strawberry from a plant in a plot to one of the categories may be affected by different random processes that are correlated. This model captures the heterogeneity between clusters (plots) and the possible unobservable heterogeneity in categorizing the plants. The result of fitting this model to the strawberry data are reported in Tables 5, 6, and 7.

The direction of the parameter estimates are similar to the results of the one random-effect models reported in Tables 2, 3, and 4. The $AIC$ statistics are estimated very close to those found in the one random-effect models. The $\chi^2$ statistics are estimated about 10 units less in the two random-effects models than in the one random-effect models, indicating that the probability of the individual categories is better estimated when two random–effects are considered. The $ICC$ is estimated significantly positive; it is more than three times the estimated value in the models with one random-effect. The standard deviations of the random-effects and the correlation are estimated significantly positive.

**4. Simulation**

In the previous section we showed that the random-effects models fit the strawberry data adequately and are much better than the homogeneous models in predicting the probability of the individual categories. In application of the random-effects models to strawberry data we do not know the true values of the parameters. Simulation study allows us to assess the performance of the random-effects models in the presence of low and high heterogeneity. In this section we report a simulation study to investigate the performance of the three models introduced in section 2 with and without a random-effects component. The simulation structure is based on the strawberry data experiment analyzed in section 3. We have designed a 3×4×4 table similar to the randomized blocked design for the strawberry data experiment and have considered 10 observations in each of the 48 clusters. We have simulated 3 types of datasets by using 3 models discussed in section 2. In each type, the 10 observations are randomly distributed into 3 categories according to the multinomial distribution with probabilities $\pi_{i1}, \pi_{i2}, \pi_{i3}$ calculated from each of the equations 1, 2, and 3. We consider the true values of the parameters to be very close to the parameter estimates in the strawberry data analysis for the calculation of the linear predictor $\delta_{ik}$ in equation 9. When the simulation of the data is completed the same model that produced the data and the other 2 models, with and without random-effects, are fitted. In order to eliminate the sampling error, and using the asymptotic property of the estimates, we replicate this procedure 100 times (simulation of the data and fitting the model to the simulated data) and report; the average of the estimates, the standard error of the estimates, the p-values, and the 95% confidence interval of the estimates. Similar to the analysis in section 3 we calculate the Pearson $\chi^2$, $C$, and $AIC$. The results from this simulation are reported in Tables 8 to 16.

The results show that the homogeneous models do not fit the simulated data well. The $\chi^2$ statistics are large with small $p-values$, even when the data are fitted by the same model that produced the data. The likelihood ratio chi-square statistic $C$ shows that all models fit the simulated data significantly different from the intercept model. The random-effects models fit the simulated data well regardless of which model has produced the simulated data. Although the direction of the parameter estimates are the same in the homogeneous and random-effects models, and also consistent with the results for the strawberry data, the difference between the $\chi^2$ statistics show that the random effects-models predict the probability of the individual categories more precisely.

We have extended the simulation study to the case that the standard error of the random-effects is 1.5. This means that we assume that the clusters are more heterogeneous. Tables 8 to 25 show that the larger the standard deviation of the random-effects, the larger the difference between the $\chi^2$ statistic for homogeneous and $\chi^2$ statistic for the random-effects models. This shows the importance of random-effects modeling.

## 5. Discussion

In this research we investigated the performance of the proportional odds, adjacent categories, and continuation ratio logit models. We introduced a random component to the linear predictor to control for heterogeneity, omitted variables, and possible positive intraclass correlation. The distribution of the random-effects is assumed to be normal. The proposed models are applied to the strawberry data (Jansen [1]). A simulation study is presented to justify the importance of random-effects modelling.

The standard error of the random-effects and the intraclass correlation are well captured in the random-effects models in the analysis of the strawberry data. We showed that the random-effects models fit the data better than homogenous models to estimate the probability of the individual categories. The likelihood ratio chi-square statistic $C$ shows that all models fit the data significantly better than the intercept model. Our analysis shows that male parent's effect is not significant and the female parent's effects are significant by using any of the three random-effects models. The intercepts are estimated negative in all models. We showed that in a two component random-effects model the probability of the individual probability categories are estimated more accurately than the one component random-effects model.

We have performed a simulation study to justify the importance of random-effects modelling. We have simulated the data consistent to the strawberry data experiment (Jansen [1]). The results show that the homogeneous models do not fit the simulated data well, even when the data are fitted by the same model that produced the data. The direction of the estimates are consistent across the models and consistent with the strawberry data analysis in the random effects-models. We extended the simulation study to the case that the standard error of the random-effects is larger. We observed the same pattern in the model estimations. Comparison of the chi-squared statistics between the random-effects and homogeneous models indicates that when the data are produced by a process that its linear predictor is affected by an unobservable random component, ignoring random-effects results in a poor fit. The simulation study shows that the larger the standard deviation of the random-effects in generating the data, the greater the importance of random-effects modelling using any of the logisitic regression models.

The simulation and application to the strawberry data confirms that random-effects modelling produces more accurate probability for the individual categories. Our analysis has not shown any preferences in using any of the three logistic regression models.

**Table 1. Data from strawberry experiment Jansen (1990)**

| Male | Female | Block | | | | | | | | | | | |
|---|---|---|---|---|---|---|---|---|---|---|---|---|---|
| | | 1 | | | 2 | | | 3 | | | 4 | | |
| | | Category | | | | | | | | | | | |
| | | 1 | 2 | 3 | 1 | 2 | 3 | 1 | 2 | 3 | 1 | 2 | 3 |
| 1 | 1 | 0 | 3 | 6 | 2 | 2 | 6 | 2 | 3 | 5 | 2 | 5 | 3 |
| 1 | 2 | 2 | 3 | 5 | 0 | 3 | 7 | 4 | 6 | 0 | 2 | 3 | 5 |
| 1 | 3 | 3 | 4 | 3 | 7 | 2 | 1 | 1 | 1 | 7 | 2 | 3 | 5 |
| 1 | 4 | 0 | 5 | 5 | 5 | 4 | 1 | 2 | 8 | 0 | 1 | 4 | 5 |
| 2 | 1 | 1 | 4 | 4 | 2 | 2 | 6 | 1 | 2 | 7 | 1 | 5 | 4 |
| 2 | 2 | 1 | 4 | 5 | 3 | 4 | 2 | 1 | 6 | 3 | 4 | 2 | 4 |
| 2 | 3 | 4 | 3 | 3 | 5 | 1 | 4 | 3 | 3 | 4 | 4 | 2 | 4 |
| 2 | 4 | 1 | 4 | 5 | 1 | 2 | 6 | 8 | 2 | 0 | 2 | 5 | 3 |
| 3 | 1 | 0 | 0 | 9 | 3 | 5 | 2 | 2 | 5 | 3 | 0 | 0 | 10 |
| 3 | 2 | 5 | 3 | 2 | 3 | 2 | 5 | 3 | 6 | 1 | 2 | 1 | 7 |
| 3 | 3 | 0 | 3 | 6 | 2 | 5 | 3 | 1 | 3 | 6 | 0 | 3 | 7 |
| 3 | 4 | 3 | 0 | 7 | 5 | 2 | 3 | 7 | 3 | 0 | 3 | 4 | 3 |

**Table 2: Proportional odds models – Strawberry data**

| Parameter | No Random Effect | | | | | Random Effect | | | | |
|---|---|---|---|---|---|---|---|---|---|---|
| | Estimate | SE[1] | P-value | LCI[2] | UCI[3] | Estimate | SE | P-value | LCI | UCI |
| $c_1$ | -2.171 | 0.287 | <.0001 | -2.738 | -1.604 | -2.388 | 0.422 | <.0001 | -3.238 | -1.539 |
| $c_2$ | -0.669 | 0.270 | 0.014 | -1.202 | -0.135 | -0.750 | 0.406 | 0.071 | -1.567 | 0.067 |
| $m_2$ | 0.117 | 0.211 | 0.581 | -0.301 | 0.535 | 0.142 | 0.321 | 0.661 | -0.504 | 0.788 |
| $m_3$ | -0.121 | 0.212 | 0.571 | -0.540 | 0.299 | -0.177 | 0.325 | 0.588 | -0.831 | 0.476 |
| $f_2$ | 0.679 | 0.248 | 0.007 | 0.188 | 1.170 | 0.789 | 0.378 | 0.042 | 0.028 | 1.549 |
| $f_3$ | 0.594 | 0.253 | 0.020 | 0.094 | 1.093 | 0.692 | 0.381 | 0.075 | -0.074 | 1.458 |
| $f_4$ | 1.015 | 0.248 | <.0001 | 0.525 | 1.506 | 1.138 | 0.379 | 0.004 | 0.375 | 1.900 |
| $b_2$ | 0.696 | 0.252 | 0.007 | 0.198 | 1.193 | 0.750 | 0.378 | 0.053 | -0.010 | 1.509 |
| $b_3$ | 0.819 | 0.245 | 0.001 | 0.334 | 1.303 | 0.869 | 0.375 | 0.025 | 0.116 | 1.622 |
| $b_4$ | 0.103 | 0.250 | 0.681 | -0.391 | 0.597 | 0.107 | 0.378 | 0.778 | -0.653 | 0.867 |
| $\sigma$ | - | - | - | - | - | 0.671 | 0.141 | <.0001 | 0.387 | 0.956 |
| $ICC^4$ | - | - | - | - | - | 0.120 | 0.045 | 0.010 | 0.031 | 0.210 |
| $\chi^2$ | 146.1 | (df = 86), p-value = 0.000 | | | | 70.0 | (df =86), p-value = 0.895 | | | |
| $C$ | 35 | (df = 8), p-value = 0.000 | | | | 16.1 | (df = 8), p-value = 0.041 | | | |
| AIC | 384.1 | | | | | 370.5 | | | | |

1. Standard error of estimate; 2. Lower 95% confidence interval; 3. Upper 95% confidence interval; 4. Intraclass correlation.

**Table 3: Adjacent categories logit models – Strawberry data**

| Parameter | No Random Effect | | | | | Random Effect | | | | |
|---|---|---|---|---|---|---|---|---|---|---|
| | Estimate | SE[1] | P-value | LCI[2] | UCI[3] | Estimate | SE | P-value | LCI | UCI |
| $c_1$ | -1.076 | 0.236 | <.0001 | -1.550 | -0.601 | -1.305 | 0.332 | 0.000 | -1.973 | -0.638 |
| $c_2$ | -0.929 | 0.204 | <.0001 | -1.339 | -0.520 | -0.964 | 0.299 | 0.002 | -1.565 | -0.362 |
| $m_2$ | 0.076 | 0.146 | 0.605 | -0.217 | 0.369 | 0.094 | 0.228 | 0.682 | -0.364 | 0.552 |
| $m_3$ | -0.068 | 0.147 | 0.644 | -0.363 | 0.227 | -0.115 | 0.231 | 0.621 | -0.580 | 0.350 |
| $f_2$ | 0.506 | 0.175 | 0.006 | 0.154 | 0.858 | 0.586 | 0.270 | 0.035 | 0.043 | 1.129 |
| $f_3$ | 0.432 | 0.176 | 0.018 | 0.079 | 0.786 | 0.490 | 0.271 | 0.076 | -0.054 | 1.034 |
| $f_4$ | 0.721 | 0.175 | 0.000 | 0.368 | 1.073 | 0.824 | 0.273 | 0.004 | 0.275 | 1.374 |
| $b_2$ | 0.480 | 0.172 | 0.008 | 0.134 | 0.826 | 0.548 | 0.267 | 0.046 | 0.010 | 1.086 |
| $b_3$ | 0.570 | 0.172 | 0.002 | 0.225 | 0.916 | 0.654 | 0.269 | 0.019 | 0.112 | 1.195 |
| $b_4$ | 0.076 | 0.174 | 0.665 | -0.275 | 0.427 | 0.094 | 0.268 | 0.729 | -0.446 | 0.634 |
| $\sigma$ | - | - | - | - | - | 0.473 | 0.104 | <.0001 | 0.264 | 0.683 |
| $ICC^4$ | - | - | - | - | - | 0.064 | 0.026 | 0.019 | 0.011 | 0.117 |
| $\chi^2$ | 145.5 (df = 86), p-value = 0.000 | | | | | 71.6 (df = 86), p-value = 0.868 | | | | |
| $C$ | 35.3 (df = 8), p-value = 0.000 | | | | | 16.8 (df = 8), p-value = 0.032 | | | | |
| $AIC$ | 383.7 | | | | | 370.5 | | | | |

1. Standard error of estimate; 2. Lower 95% confidence interval; 3. Upper 95% confidence interval; 4. Intraclass correlation.

**Table 4: Continuation ratio logit models – Strawberry data**

| Parameter | No Random Effect | | | | | Random Effect | | | | |
|---|---|---|---|---|---|---|---|---|---|---|
| | Estimate | SE[1] | P-value | LCI[2] | UCI[3] | Estimate | SE | P-value | LCI | UCI |
| $c_1$ | -2.021 | 0.265 | <.0001 | -2.545 | -1.498 | -2.221 | 0.390 | <.0001 | -3.006 | -1.436 |
| $c_2$ | -1.068 | 0.250 | <.0001 | -1.563 | -0.574 | -1.106 | 0.375 | 0.005 | -1.860 | -0.352 |
| $m_2$ | 0.062 | 0.188 | 0.743 | -0.310 | 0.433 | 0.078 | 0.292 | 0.792 | -0.510 | 0.665 |
| $m_3$ | -0.127 | 0.190 | 0.505 | -0.502 | 0.248 | -0.189 | 0.297 | 0.527 | -0.786 | 0.408 |
| $f_2$ | 0.613 | 0.223 | 0.007 | 0.173 | 1.054 | 0.709 | 0.345 | 0.045 | 0.016 | 1.403 |
| $f_3$ | 0.478 | 0.226 | 0.036 | 0.031 | 0.925 | 0.571 | 0.347 | 0.106 | -0.127 | 1.268 |
| $f_4$ | 0.920 | 0.223 | <.0001 | 0.479 | 1.361 | 1.048 | 0.347 | 0.004 | 0.350 | 1.746 |
| $b_2$ | 0.568 | 0.223 | 0.012 | 0.127 | 1.008 | 0.640 | 0.343 | 0.068 | -0.050 | 1.330 |
| $b_3$ | 0.777 | 0.220 | 0.001 | 0.343 | 1.212 | 0.859 | 0.343 | 0.016 | 0.170 | 1.548 |
| $b_4$ | 0.091 | 0.225 | 0.686 | -0.353 | 0.535 | 0.088 | 0.344 | 0.799 | -0.604 | 0.780 |
| $\sigma$ | - | - | - | - | - | 0.616 | 0.128 | <.0001 | 0.358 | 0.874 |
| $ICC^4$ | - | - | - | - | - | 0.104 | 0.039 | 0.010 | 0.026 | 0.181 |
| $\chi^2$ | 147.1 (df = 86), p-value = 0.000 | | | | | 68.6 (df = 86), p-value = 0.916 | | | | |
| $C$ | 35.4 (df = 8), p-value = 0.000 | | | | | 16.7 (df = 8), p-value = 0.030 | | | | |
| $AIC$ | 383.6 | | | | | 368.9 | | | | |

1. Standard error of estimate; 2. Lower 95% confidence interval; 3. Upper 95% confidence interval; 4. Intraclass correlation.

**Table 5: Proportional odds model with two random effects – Strawberry data**

| Parameter | Estimate | SE[1] | P-value | LCI[2] | UCI[3] |
|---|---|---|---|---|---|
| $c_1$ | -2.427 | 0.428 | <.0001 | -3.289 | -1.565 |
| $c_2$ | -0.797 | 0.417 | 0.062 | -1.637 | 0.042 |
| $m_2$ | 0.178 | 0.328 | 0.591 | -0.483 | 0.838 |
| $m_3$ | -0.121 | 0.337 | 0.723 | -0.799 | 0.558 |
| $f_2$ | 0.801 | 0.381 | 0.041 | 0.034 | 1.569 |
| $f_3$ | 0.751 | 0.396 | 0.064 | -0.046 | 1.548 |
| $f_4$ | 1.143 | 0.381 | 0.004 | 0.376 | 1.909 |
| $b_2$ | 0.803 | 0.385 | 0.043 | 0.027 | 1.579 |
| $b_3$ | 0.840 | 0.384 | 0.034 | 0.066 | 1.614 |
| $b_4$ | 0.118 | 0.382 | 0.759 | -0.650 | 0.887 |
| $\sigma_1$ | 0.609 | 0.185 | 0.002 | 0.235 | 0.982 |
| $\sigma_2$ | 0.738 | 0.171 | <.0001 | 0.393 | 1.083 |
| $\rho$ | 0.933 | 0.165 | <.0001 | 0.602 | 1.000 |
| ICC[4] | 0.348 | 0.098 | 0.001 | 0.151 | 0.544 |
| $\chi^2$ | 61.3 1 | (df = 86), p-value = 0.980 | | | |
| $C$ | 16.1 | (df = 8), p-value = 0.041 | | | |
| $AIC$ | 373.8 | | | | |

1. Standard error of estimate; 2. Lower 95% confidence interval; 3. Upper 95% confidence interval; 4. Intraclass correlation.

**Table 6: Adjacent categories logit model with two random effects – Strawberry data**

| Parameter | Estimate | SE[1] | P-value | LCI[2] | UCI[3] |
|---|---|---|---|---|---|
| $c_1$ | -1.293 | 0.335 | 0.000 | -1.967 | -0.620 |
| $c_2$ | -1.006 | 0.310 | 0.002 | -1.630 | -0.382 |
| $m_2$ | 0.116 | 0.229 | 0.614 | -0.344 | 0.577 |
| $m_3$ | -0.065 | 0.237 | 0.786 | -0.542 | 0.412 |
| $f_2$ | 0.587 | 0.271 | 0.036 | 0.041 | 1.133 |
| $f_3$ | 0.532 | 0.278 | 0.061 | -0.027 | 1.091 |
| $f_4$ | 0.822 | 0.272 | 0.004 | 0.273 | 1.370 |
| $b_2$ | 0.577 | 0.268 | 0.036 | 0.038 | 1.116 |
| $b_3$ | 0.623 | 0.273 | 0.027 | 0.073 | 1.173 |
| $b_4$ | 0.098 | 0.270 | 0.717 | -0.444 | 0.641 |
| $\sigma_1$ | 0.404 | 0.255 | 0.120 | -0.109 | 0.917 |
| $\sigma_2$ | 0.652 | 0.198 | 0.002 | 0.253 | 1.051 |
| $\rho$ | 0.493 | 0.770 | 0.525 | -1.057 | 2.043 |
| ICC[4] | 0.205 | 0.074 | 0.008 | 0.056 | 0.354 |
| $\chi^2$ | 60.1 | (df = 86), p-value = 0.985 | | | |
| $C$ | 16.5 | (df = 8), p-value = 0.036 | | | |
| $AIC$ | 373.4 | | | | |

1. Standard error of estimate; 2. Lower 95% confidence interval; 3. Upper 95% confidence interval; 4. Intraclass correlation.

**Table 7: Continuation ratio logit model with two random effects – Strawberry data**

| Parameter | Estimate | SE[1] | P-value | LCI[2] | UCI[3] |
|---|---|---|---|---|---|
| $c_1$ | -2.230 | 0.400 | <.0001 | -3.034 | -1.425 |
| $c_2$ | -1.115 | 0.390 | 0.006 | -1.900 | -0.329 |
| $m_2$ | 0.074 | 0.299 | 0.805 | -0.527 | 0.675 |
| $m_3$ | -0.190 | 0.305 | 0.537 | -0.805 | 0.425 |
| $f_2$ | 0.713 | 0.348 | 0.046 | 0.013 | 1.413 |
| $f_3$ | 0.568 | 0.364 | 0.126 | -0.165 | 1.301 |
| $f_4$ | 1.049 | 0.350 | 0.004 | 0.345 | 1.753 |
| $b_2$ | 0.642 | 0.355 | 0.077 | -0.073 | 1.358 |
| $b_3$ | 0.868 | 0.347 | 0.016 | 0.170 | 1.565 |
| $b_4$ | 0.091 | 0.345 | 0.793 | -0.604 | 0.786 |
| $\sigma_1$ | 0.634 | 0.191 | 0.002 | 0.249 | 1.020 |
| $\sigma_2$ | 0.636 | 0.205 | 0.003 | 0.223 | 1.050 |
| $\rho$ | 0.884 | 0.346 | 0.014 | 0.188 | 1.000 |
| ICC[4] | 0.316 | 0.090 | 0.001 | 0.134 | 0.498 |
| $\chi^2$ | 62.3 1 | (df = 86), p-value = 0.975 | | | |
| $C$ | 16.7 | (df = 8), p-value = 0.033 | | | |
| $AIC$ | 372.8 | | | | |

1. Standard error of estimate; 2. Lower 95% confidence interval; 3. Upper 95% confidence interval; 4. Intraclass correlation

**Table 8: Simulation result from data produced by proportional odds random-effects model with ($\sigma = 0.6$) and fitted by proportional odds model**

| Parameter | No Random Effect | | | | Random Effect | | | |
|---|---|---|---|---|---|---|---|---|
| | Estimate | SD[1] | LCI[2] | UCI[3] | Estimate | SD | LCI | UCI |
| $c_1(-2^5)$ | -1.754 | 0.378 | -1.828 | -1.680 | -1.806 | 0.386 | -1.882 | -1.730 |
| $c_2(-1)$ | -1.072 | 0.374 | -1.145 | -0.998 | -1.103 | 0.382 | -1.178 | -1.028 |
| $m_2(0.1)$ | 0.068 | 0.280 | 0.013 | 0.123 | 0.071 | 0.289 | 0.014 | 0.127 |
| $m_3(-0.2)$ | -0.204 | 0.293 | -0.261 | -0.146 | -0.210 | 0.303 | -0.270 | -0.151 |
| $f_2(0.7)$ | 0.613 | 0.341 | 0.547 | 0.680 | 0.633 | 0.349 | 0.564 | 0.701 |
| $f_3(0.6)$ | 0.481 | 0.345 | 0.414 | 0.549 | 0.494 | 0.354 | 0.425 | 0.563 |
| $f_4(1)$ | 0.846 | 0.358 | 0.776 | 0.916 | 0.871 | 0.365 | 0.800 | 0.943 |
| $b_2(0.6)$ | 0.470 | 0.360 | 0.399 | 0.540 | 0.485 | 0.367 | 0.413 | 0.557 |
| $b_3(0.1)$ | 0.709 | 0.357 | 0.639 | 0.779 | 0.732 | 0.363 | 0.660 | 0.803 |
| $b_4(0.9)$ | 0.083 | 0.368 | 0.011 | 0.155 | 0.087 | 0.381 | 0.012 | 0.161 |
| $\sigma(0.6)$ | - | - | - | - | 0.287 | 0.244 | 0.239 | 0.335 |
| $ICC^4$ | - | - | - | - | 0.115 | 0.087 | 0.097 | 0.132 |
| $\chi^2$ | 110.2 | (df = 86), p-value = 0.040 | | | 83.0 | (df =86), p-value = 0.572 | | |
| $C$ | 34 | (df = 8), p-value = 0.000 | | | 22.2 | (df = 8), p-value = 0.004 | | |
| AIC | 384.1 | | | | 370.5 | | | |

1. Standard deviation of estimate; 2. Lower 95% confidence interval; 3. Upper 95% confidence interval; 4. Intraclass correlation.
5. True value of the parameter.

**Table 9: Simulation result from data produced by proportional odds random-effects model with ($\sigma = 0.6$) and fitted by adjacent categories logit model**

| Parameter | No Random Effect | | | | Random Effect | | | |
|---|---|---|---|---|---|---|---|---|
| | Estimate | SD[1] | LCI[2] | UCI[3] | Estimate | SD | LCI | UCI |
| $c_1(-2^5)$ | 0.133 | 0.275 | 0.079 | 0.187 | 0.079 | 0.286 | 0.023 | 0.135 |
| $c_2(-1)$ | -1.823 | 0.260 | -1.874 | -1.772 | -1.830 | 0.266 | -1.882 | -1.778 |
| $m_2(0.1)$ | 0.041 | 0.167 | 0.009 | 0.074 | 0.044 | 0.173 | 0.010 | 0.078 |
| $m_3(-0.2)$ | -0.120 | 0.176 | -0.155 | -0.086 | -0.125 | 0.182 | -0.160 | -0.089 |
| $f_2(0.7)$ | 0.372 | 0.203 | 0.333 | 0.412 | 0.385 | 0.209 | 0.344 | 0.426 |
| $f_3(0.6)$ | 0.295 | 0.209 | 0.254 | 0.336 | 0.304 | 0.215 | 0.261 | 0.346 |
| $f_4(1)$ | 0.510 | 0.208 | 0.469 | 0.550 | 0.527 | 0.215 | 0.485 | 0.569 |
| $b_2(0.6)$ | 0.281 | 0.215 | 0.238 | 0.323 | 0.291 | 0.220 | 0.248 | 0.334 |
| $b_3(0.1)$ | 0.420 | 0.212 | 0.378 | 0.461 | 0.435 | 0.218 | 0.392 | 0.478 |
| $b_4(0.9)$ | 0.048 | 0.220 | 0.005 | 0.092 | 0.051 | 0.229 | 0.006 | 0.096 |
| $\sigma(0.6)$ | - | - | - | - | 0.053 | 0.220 | 0.010 | 0.096 |
| $ICC^4$ | - | - | - | - | 0.047 | 0.037 | 0.039 | 0.054 |
| $\chi^2$ | 109.6 | (df = 86), p-value = 0.044 | | | 82.6 | (df =86), p-value = 0.584 | | |
| $C$ | 35.4 | (df = 8), p-value = 0.000 | | | 22.4 | (df = 8), p-value = 0.004 | | |
| AIC | 330.4 | | | | 329.6 | | | |

1. Standard deviation of estimate; 2. Lower 95% confidence interval; 3. Upper 95% confidence interval; 4. Intraclass correlation.
5. True value of the parameter.

Table 10: Simulation result from data produced by proportional odds random-effects model with ($\sigma = 0.6$) and fitted by continuation ratio logit model

| Parameter | No Random Effect | | | | Random Effect | | | |
|---|---|---|---|---|---|---|---|---|
| | Estimate | SD[1] | LCI[2] | UCI[3] | Estimate | SD | LCI | UCI |
| $c_1(-2^5)$ | -1.646 | 0.339 | -1.713 | -1.580 | -1.685 | 0.345 | -1.753 | -1.618 |
| $c_2(-1)$ | -2.031 | 0.363 | -2.102 | -1.960 | -2.045 | 0.367 | -2.117 | -1.973 |
| $m_2(0.1)$ | 0.057 | 0.247 | 0.009 | 0.106 | 0.060 | 0.254 | 0.010 | 0.110 |
| $m_3(-0.2)$ | -0.179 | 0.260 | -0.229 | -0.128 | -0.183 | 0.267 | -0.235 | -0.131 |
| $f_2(0.7)$ | 0.536 | 0.302 | 0.477 | 0.595 | 0.550 | 0.308 | 0.490 | 0.610 |
| $f_3(0.6)$ | 0.419 | 0.312 | 0.358 | 0.481 | 0.429 | 0.317 | 0.367 | 0.491 |
| $f_4(1)$ | 0.728 | 0.318 | 0.666 | 0.791 | 0.747 | 0.323 | 0.684 | 0.810 |
| $b_2(0.6)$ | 0.411 | 0.322 | 0.348 | 0.474 | 0.423 | 0.328 | 0.358 | 0.487 |
| $b_3(0.1)$ | 0.608 | 0.313 | 0.547 | 0.669 | 0.624 | 0.318 | 0.562 | 0.686 |
| $b_4(0.9)$ | 0.074 | 0.336 | 0.009 | 0.140 | 0.076 | 0.346 | 0.009 | 0.144 |
| $\sigma(0.6)$ | - | - | - | - | 0.226 | 0.217 | 0.183 | 0.268 |
| ICC[4] | - | - | - | - | 0.131 | 0.060 | 0.119 | 0.143 |
| $\chi^2$ | 112.6 | (df = 86), p-value = 0.029 | | | 89.2 | (df =86), p-value = 0.385 | | |
| C | 32.0 | (df = 8), p-value = 0.000 | | | 21.5 | (df = 8), p-value = 0.006 | | |
| AIC | 333.8 | | | | 333.6 | | | |

1. Standard deviation of estimate; 2. Lower 95% confidence interval; 3. Upper 95% confidence interval; 4. Intraclass correlation.
5. True value of the parameter.

Table 11: Simulation result from data produced by adjacent categories logit random-effects model with ($\sigma = 0.6$) and fitted by proportional odds model

| Parameter | No Random Effect | | | | Random Effect | | | |
|---|---|---|---|---|---|---|---|---|
| | Estimate | SD[1] | LCI[2] | UCI[3] | Estimate | SD | LCI | UCI |
| $c_1(-2^5)$ | -2.409 | 0.354 | 0.125 | -2.478 | -2.520 | 0.382 | -2.595 | -2.445 |
| $c_2(-1)$ | -0.888 | 0.346 | 0.128 | -0.956 | -0.930 | 0.369 | -1.002 | -0.858 |
| $m_2(0.1)$ | 0.113 | 0.302 | 0.092 | 0.053 | 0.116 | 0.317 | 0.054 | 0.178 |
| $m_3(-0.2)$ | -0.185 | 0.258 | 0.067 | -0.235 | -0.197 | 0.269 | -0.249 | -0.144 |
| $f_2(0.7)$ | 0.593 | 0.345 | 0.162 | 0.526 | 0.625 | 0.365 | 0.553 | 0.697 |
| $f_3(0.6)$ | 0.516 | 0.336 | 0.146 | 0.450 | 0.544 | 0.352 | 0.475 | 0.613 |
| $f_4(1)$ | 0.919 | 0.362 | 0.164 | 0.848 | 0.967 | 0.381 | 0.892 | 1.041 |
| $b_2(0.6)$ | 0.520 | 0.304 | 0.125 | 0.460 | 0.550 | 0.323 | 0.486 | 0.613 |
| $b_3(0.1)$ | 0.817 | 0.322 | 0.110 | 0.754 | 0.861 | 0.343 | 0.794 | 0.928 |
| $b_4(0.9)$ | 0.074 | 0.330 | 0.110 | 0.009 | 0.081 | 0.349 | 0.013 | 0.149 |
| $\sigma(0.6)$ | - | - | - | - | 0.438 | 0.190 | 0.401 | 0.476 |
| ICC[4] | - | - | - | - | 0.176 | 0.092 | 0.157 | 0.193 |
| $\chi^2$ | 127.8 | (df = 86), p-value = 0.002 | | | 82.4 | (df =86), p-value = 0.590 | | |
| C | 40.9 | (df = 8), p-value = 0.000 | | | 23.2 | (df = 8), p-value = 0.003 | | |
| AIC | 352.6 | | | | 348.6 | | | |

1. Standard deviation of estimate; 2. Lower 95% confidence interval; 3. Upper 95% confidence interval; 4. Intraclass correlation.
5. True value of the parameter.

**Table 12: Simulation result from data produced by adjacent categories logit random-effects model with ($\sigma = 0.6$) and fitted by adjacent categories logit model**

| Parameter | No Random Effect | | | | Random Effect | | | |
|---|---|---|---|---|---|---|---|---|
| | Estimate | SD[1] | LCI[2] | UCI[3] | Estimate | SD | LCI | UCI |
| $c_1(-2^5)$ | -1.229 | 0.299 | -1.288 | -1.170 | -1.357 | 0.321 | -1.420 | -1.294 |
| $c_2(-1)$ | -1.134 | 0.276 | -1.188 | -1.080 | -1.154 | 0.292 | -1.211 | -1.097 |
| $m_2(0.1)$ | 0.076 | 0.213 | 0.034 | 0.118 | 0.082 | 0.225 | 0.037 | 0.126 |
| $m_3(-0.2)$ | -0.136 | 0.183 | -0.172 | -0.100 | -0.144 | 0.192 | -0.182 | -0.107 |
| $f_2(0.7)$ | 0.446 | 0.255 | 0.396 | 0.496 | 0.472 | 0.270 | 0.419 | 0.525 |
| $f_3(0.6)$ | 0.389 | 0.251 | 0.340 | 0.438 | 0.413 | 0.265 | 0.361 | 0.464 |
| $f_4(1)$ | 0.672 | 0.257 | 0.622 | 0.723 | 0.713 | 0.272 | 0.659 | 0.766 |
| $b_2(0.6)$ | 0.378 | 0.214 | 0.336 | 0.420 | 0.402 | 0.228 | 0.357 | 0.446 |
| $b_3(0.1)$ | 0.585 | 0.226 | 0.540 | 0.629 | 0.622 | 0.244 | 0.574 | 0.669 |
| $b_4(0.9)$ | 0.051 | 0.244 | 0.003 | 0.099 | 0.057 | 0.262 | 0.006 | 0.109 |
| $\sigma(0.6)$ | - | - | - | - | 0.224 | 0.260 | 0.173 | 0.275 |
| $ICC^4$ | - | - | - | - | 0.017 | 0.043 | -0.070 | 0.102 |
| $\chi^2$ | 126.7 | (df = 86), p-value = 0.003 | | | 81.0 | (df =86), p-value = 0.632 | | |
| $C$ | 43.6 | (df = 8), p-value = 0.000 | | | 23.7 | (df = 8), p-value = 0.003 | | |
| $AIC$ | 349.9 | | | | 345.7 | | | |

1. Standard deviation of estimate; 2. Lower 95% confidence interval; 3. Upper 95% confidence interval; 4. Intraclass correlation.
5. True value of the parameter.

**Table 13: Simulation result from data produced by adjacent categories logit random-effects model with ($\sigma = 0.6$) and fitted by continuation ratio logit model**

| Parameter | No Random Effect | | | | Random Effect | | | |
|---|---|---|---|---|---|---|---|---|
| | Estimate | SD[1] | LCI[2] | UCI[3] | Estimate | SD | LCI | UCI |
| $c_1(-2^5)$ | -2.291 | 0.321 | -2.354 | -2.228 | -2.382 | 0.346 | -2.449 | -2.314 |
| $c_2(-1)$ | -1.238 | 0.320 | -1.301 | -1.176 | -1.258 | 0.336 | -1.324 | -1.192 |
| $m_2(0.1)$ | 0.100 | 0.261 | 0.049 | 0.151 | 0.103 | 0.275 | 0.049 | 0.157 |
| $m_3(-0.2)$ | -0.158 | 0.228 | -0.203 | -0.114 | -0.167 | 0.238 | -0.214 | -0.121 |
| $f_2(0.7)$ | 0.515 | 0.311 | 0.454 | 0.576 | 0.542 | 0.325 | 0.478 | 0.606 |
| $f_3(0.6)$ | 0.451 | 0.303 | 0.392 | 0.510 | 0.475 | 0.316 | 0.413 | 0.537 |
| $f_4(1)$ | 0.789 | 0.322 | 0.726 | 0.852 | 0.829 | 0.337 | 0.763 | 0.895 |
| $b_2(0.6)$ | 0.453 | 0.270 | 0.400 | 0.506 | 0.477 | 0.285 | 0.421 | 0.533 |
| $b_3(0.1)$ | 0.696 | 0.278 | 0.641 | 0.750 | 0.733 | 0.295 | 0.676 | 0.791 |
| $b_4(0.9)$ | 0.067 | 0.301 | 0.008 | 0.126 | 0.073 | 0.316 | 0.011 | 0.135 |
| $\sigma(0.6)$ | - | - | - | - | 0.362 | 0.183 | 0.326 | 0.398 |
| $ICC^4$ | - | - | - | - | 0.134 | 0.078 | 0.119 | 0.149 |
| $\chi^2$ | 130.7 | (df = 86), p-value = 0.001 | | | 89.7 | (df =86), p-value = 0.371 | | |
| $C$ | 37.7 | (df = 8), p-value = 0.000 | | | 22.5 | (df = 8), p-value = 0.004 | | |
| $AIC$ | 355.8 | | | | 352.9 | | | |

1. Standard deviation of estimate; 2. Lower 95% confidence interval; 3. Upper 95% confidence interval; 4. Intraclass correlation.
5. True value of the parameter.

**Table 14: Simulation result from data produced by continuation ratio logit random-effects model with ($\sigma = 0.6$) and fitted by proportional odds model**

| Parameter | No Random Effect | | | | Random Effect | | | |
|---|---|---|---|---|---|---|---|---|
| | Estimate | SD[1] | LCI[2] | UCI[3] | Estimate | SD | LCI | UCI |
| $c_1$(-2[5]) | -1.752 | 0.359 | -1.822 | -1.681 | -1.807 | 0.372 | -1.880 | -1.734 |
| $c_2$(-1) | -0.707 | 0.353 | -0.776 | -0.637 | -0.729 | 0.363 | -0.800 | -0.658 |
| $m_2$(0.1) | 0.060 | 0.275 | 0.006 | 0.114 | 0.062 | 0.285 | 0.006 | 0.118 |
| $m_3$(-0.2) | -0.196 | 0.276 | -0.250 | -0.142 | -0.202 | 0.287 | -0.258 | -0.146 |
| $f_2$(0.7) | 0.612 | 0.331 | 0.547 | 0.677 | 0.633 | 0.341 | 0.566 | 0.700 |
| $f_3$(0.6) | 0.486 | 0.331 | 0.421 | 0.551 | 0.501 | 0.341 | 0.435 | 0.568 |
| $f_4$(1) | 0.843 | 0.345 | 0.776 | 0.911 | 0.872 | 0.355 | 0.802 | 0.941 |
| $b_2$(0.6) | 0.471 | 0.348 | 0.403 | 0.539 | 0.488 | 0.357 | 0.418 | 0.558 |
| $b_3$(0.1) | 0.703 | 0.344 | 0.635 | 0.770 | 0.727 | 0.353 | 0.658 | 0.796 |
| $b_4$(0.9) | 0.087 | 0.358 | 0.016 | 0.157 | 0.092 | 0.372 | 0.019 | 0.164 |
| $\sigma$(0.6) | - | - | - | - | 0.274 | 0.263 | 0.223 | 0.325 |
| $ICC$[4] | - | - | - | - | 0.117 | 0.086 | 0.100 | 0.133 |
| $\chi^2$ | 111.3 | (df = 86), p-value = 0.035 | | | 81.6 | (df =86), p-value = 0.610 | | |
| $C$ | 36.5 | (df = 8), p-value = 0.000 | | | 23.0 | (df = 8), p-value = 0.034 | | |
| AIC | 349.3 | | | | 348.1 | | | |

1. Standard deviation of estimate; 2. Lower 95% confidence interval; 3. Upper 95% confidence interval; 4. Intraclass correlation.
5. True value of the parameter.

**Table 15: Simulation result from data produced by continuation ratio logit random-effects model with ($\sigma = 0.6$) and fitted by adjacent categories logit model**

| Parameter | No Random Effect | | | | Random Effect | | | |
|---|---|---|---|---|---|---|---|---|
| | Estimate | SD[1] | LCI[2] | UCI[3] | Estimate | SD | LCI | UCI |
| $c_1$(-2[5]) | -0.359 | 0.264 | -0.410 | -0.307 | -0.417 | 0.280 | -0.472 | -0.362 |
| $c_2$(-1) | -1.218 | 0.251 | -1.268 | -1.169 | -1.222 | 0.259 | -1.273 | -1.171 |
| $m_2$(0.1) | 0.040 | 0.174 | 0.006 | 0.075 | 0.042 | 0.182 | 0.006 | 0.078 |
| $m_3$(-0.2) | -0.124 | 0.178 | -0.159 | -0.089 | -0.129 | 0.185 | -0.165 | -0.093 |
| $f_2$(0.7) | 0.396 | 0.211 | 0.354 | 0.437 | 0.411 | 0.219 | 0.368 | 0.454 |
| $f_3$(0.6) | 0.316 | 0.215 | 0.274 | 0.359 | 0.327 | 0.223 | 0.283 | 0.371 |
| $f_4$(1) | 0.540 | 0.217 | 0.498 | 0.583 | 0.561 | 0.225 | 0.517 | 0.605 |
| $b_2$(0.6) | 0.300 | 0.223 | 0.256 | 0.344 | 0.313 | 0.230 | 0.268 | 0.358 |
| $b_3$(0.1) | 0.443 | 0.219 | 0.400 | 0.486 | 0.461 | 0.227 | 0.416 | 0.505 |
| $b_4$(0.9) | 0.053 | 0.230 | 0.008 | 0.098 | 0.057 | 0.240 | 0.010 | 0.104 |
| $\sigma$(0.6) | - | - | - | - | 0.109 | 0.217 | 0.066 | 0.151 |
| $ICC$[4] | - | - | - | - | 0.053 | 0.041 | 0.045 | 0.061 |
| $\chi^2$ | 111.0 | (df = 86), p-value = 0.036 | | | 82.1 | (df =86), p-value = 0.599 | | |
| $C$ | 37.0 | (df = 8), p-value = 0.000 | | | 23.1 | (df = 8), p-value = 0.003 | | |
| AIC | 348.8 | | | | 347.6 | | | |

1. Standard deviation of estimate; 2. Lower 95% confidence interval; 3. Upper 95% confidence interval; 4. Intraclass correlation.
5. True value of the parameter.

**Table 16: Simulation result from data produced by continuation ratio logit random-effects model with ($\sigma = 0.6$) and fitted by continuation ratio logit model**

| Parameter | No Random Effect | | | | Random Effect | | | |
|---|---|---|---|---|---|---|---|---|
| | Estimate | SD[1] | LCI[2] | UCI[3] | Estimate | SD | LCI | UCI |
| $c_1(-2^5)$ | -1.631 | 0.317 | -1.693 | -1.569 | -1.673 | 0.328 | -1.737 | -1.608 |
| $c_2(-1)$ | -1.372 | 0.326 | -1.435 | -1.308 | -1.377 | 0.333 | -1.443 | -1.312 |
| $m_2(0.1)$ | 0.049 | 0.237 | 0.002 | 0.095 | 0.050 | 0.245 | 0.002 | 0.098 |
| $m_3(-0.2)$ | -0.167 | 0.240 | -0.214 | -0.120 | -0.171 | 0.247 | -0.220 | -0.123 |
| $f_2(0.7)$ | 0.524 | 0.289 | 0.467 | 0.581 | 0.540 | 0.296 | 0.482 | 0.598 |
| $f_3(0.6)$ | 0.416 | 0.293 | 0.359 | 0.474 | 0.428 | 0.298 | 0.370 | 0.487 |
| $f_4(1)$ | 0.713 | 0.300 | 0.654 | 0.771 | 0.735 | 0.306 | 0.675 | 0.795 |
| $b_2(0.6)$ | 0.404 | 0.305 | 0.345 | 0.464 | 0.418 | 0.312 | 0.357 | 0.480 |
| $b_3(0.1)$ | 0.590 | 0.297 | 0.532 | 0.648 | 0.608 | 0.303 | 0.549 | 0.668 |
| $b_4(0.9)$ | 0.075 | 0.320 | 0.012 | 0.138 | 0.078 | 0.331 | 0.013 | 0.143 |
| $\sigma(0.6)$ | - | - | - | - | 0.244 | 0.192 | 0.207 | 0.282 |
| $ICC^4$ | - | - | - | - | 0.082 | 0.069 | 0.069 | 0.096 |
| $\chi^2$ | 113.5 | (df = 86), p-value = 0.029 | | | 87.8 | (df =86), p-value = 0.426 | | |
| $C$ | 33.7 | (df = 8), p-value = 0.000 | | | 22.3 | (df = 8), p-value = 0.004 | | |
| $AIC$ | 352.1 | | | | 351.5 | | | |

1. Standard deviation of estimate; 2. Lower 95% confidence interval; 3. Upper 95% confidence interval; 4. Intraclass correlation.
5. True value of the parameter.

**Table 17: Simulation result from data produced by proportional odds random-effects model with ($\sigma = 1.5$) and fitted by proportional odds model**

| Parameter | No Random Effect | | | | Random Effect | | | |
|---|---|---|---|---|---|---|---|---|
| | Estimate | SD[1] | LCI[2] | UCI[3] | Estimate | SD | LCI | UCI |
| $c_1(-2^5)$ | -1.430 | 0.484 | -1.524 | -1.335 | -1.749 | 0.582 | -1.864 | -1.635 |
| $c_2(-1)$ | -0.933 | 0.472 | -1.025 | -0.841 | -1.134 | 0.569 | -1.245 | -1.022 |
| $m_2(0.1)$ | 0.081 | 0.413 | 0.000 | 0.162 | 0.111 | 0.513 | 0.011 | 0.212 |
| $m_3(-0.2)$ | -0.137 | 0.406 | -0.216 | -0.057 | -0.183 | 0.492 | -0.279 | -0.086 |
| $f_2(0.7)$ | 0.531 | 0.592 | 0.415 | 0.647 | 0.652 | 0.735 | 0.508 | 0.796 |
| $f_3(0.6)$ | 0.494 | 0.517 | 0.393 | 0.596 | 0.601 | 0.642 | 0.475 | 0.727 |
| $f_4(1)$ | 0.730 | 0.490 | 0.634 | 0.826 | 0.903 | 0.609 | 0.783 | 1.022 |
| $b_2(0.6)$ | 0.399 | 0.460 | 0.309 | 0.489 | 0.497 | 0.581 | 0.383 | 0.610 |
| $b_3(0.1)$ | 0.527 | 0.461 | 0.436 | 0.617 | 0.654 | 0.570 | 0.542 | 0.765 |
| $b_4(0.9)$ | 0.002 | 0.523 | -0.101 | 0.104 | 0.002 | 0.651 | -0.126 | 0.130 |
| $\sigma(1.5)$ | - | - | - | - | 1.135 | 0.206 | 1.094 | 1.175 |
| $ICC^4$ | - | - | - | - | 0.553 | 0.088 | 0.536 | 0.570 |
| $\chi^2$ | 198.6 | (df = 86), p-value = 0.000 | | | 60.5 | (df =86), p-value = 0.985 | | |
| $C$ | 41.5 | (df = 8), p-value = 0.000 | | | 13.3 | (df = 8), p-value = 0.102 | | |
| $AIC$ | 396.1 | | | | 347.7 | | | |

1. Standard deviation of estimate; 2. Lower 95% confidence interval; 3. Upper 95% confidence interval; 4. Intraclass correlation.
5. True value of the parameter.

**Table 18: Simulation result from data produced by proportional odds random-effects model with ($\sigma = 1.5$) by adjacent categories logit model**

| Parameter | No Random Effect | | | | Random Effect | | | |
|---|---|---|---|---|---|---|---|---|
| | Estimate | SD[1] | LCI[2] | UCI[3] | Estimate | SD | LCI | UCI |
| $c_1(-2^5)$ | 0.670 | 0.348 | 0.602 | 0.739 | 0.316 | 0.405 | 0.236 | 0.395 |
| $c_2(-1)$ | -2.020 | 0.304 | -2.079 | -1.960 | -2.034 | 0.353 | -2.103 | -1.965 |
| $m_2(0.1)$ | 0.046 | 0.237 | -0.001 | 0.092 | 0.066 | 0.301 | 0.007 | 0.125 |
| $m_3(-0.2)$ | -0.078 | 0.231 | -0.123 | -0.033 | -0.105 | 0.285 | -0.161 | -0.049 |
| $f_3(0.6)$ | 0.305 | 0.337 | 0.239 | 0.371 | 0.384 | 0.428 | 0.301 | 0.468 |
| $f_4(1)$ | 0.283 | 0.296 | 0.225 | 0.341 | 0.355 | 0.376 | 0.281 | 0.428 |
| $b_2(0.6)$ | 0.417 | 0.281 | 0.362 | 0.472 | 0.529 | 0.357 | 0.459 | 0.599 |
| $b_3(0.1)$ | 0.227 | 0.263 | 0.176 | 0.278 | 0.288 | 0.338 | 0.221 | 0.354 |
| $b_4(0.9)$ | 0.297 | 0.264 | 0.246 | 0.349 | 0.378 | 0.335 | 0.312 | 0.443 |
| $\sigma(1.5)$ | - | - | - | - | 0.000 | 0.381 | -0.075 | 0.075 |
| $ICC^4$ | - | - | - | - | 0.301 | 0.077 | 0.2861 | 0.316 |
| $\chi^2$ | 198.0 | (df = 86), p-value = 0.000 | | | 64.0 | (df =86), p-value = 0.964 | | |
| $C$ | 42.0 | (df = 8), p-value = 0.000 | | | 13.3 | (df = 8), p-value = 0.102 | | |
| AIC | 395.6 | | | | 347.4 | | | |

1. Standard deviation of estimate; 2. Lower 95% confidence interval; 3. Upper 95% confidence interval; 4. Intraclass correlation.
5. True value of the parameter.

**Table 19: Simulation result from data produced by proportional odds random-effects model with ($\sigma = 1.5$) by continuation ratio logit model**

| Parameter | No Random Effect | | | | Random Effect | | | |
|---|---|---|---|---|---|---|---|---|
| | Estimate | SD[1] | LCI[2] | UCI[3] | Estimate | SD | LCI | UCI |
| $c_1(-2^5)$ | -1.347 | 0.438 | -1.432 | -1.261 | -1.620 | 0.519 | -1.722 | -1.519 |
| $c_2(-1)$ | -2.237 | 0.438 | -2.323 | -2.151 | -2.299 | 0.510 | -2.399 | -2.199 |
| $m_2(0.1)$ | 0.071 | 0.362 | 0.000 | 0.141 | 0.102 | 0.445 | 0.014 | 0.189 |
| $m_3(-0.2)$ | -0.123 | 0.357 | -0.193 | -0.053 | -0.155 | 0.427 | -0.239 | -0.072 |
| $f_2(0.7)$ | 0.468 | 0.534 | 0.363 | 0.573 | 0.566 | 0.645 | 0.440 | 0.692 |
| $f_3(0.6)$ | 0.443 | 0.460 | 0.353 | 0.533 | 0.527 | 0.562 | 0.417 | 0.637 |
| $f_4(1)$ | 0.646 | 0.439 | 0.559 | 0.732 | 0.782 | 0.532 | 0.678 | 0.887 |
| $b_2(0.6)$ | 0.352 | 0.417 | 0.270 | 0.434 | 0.434 | 0.513 | 0.334 | 0.534 |
| $b_3(0.1)$ | 0.458 | 0.407 | 0.378 | 0.537 | 0.564 | 0.496 | 0.467 | 0.661 |
| $b_4(0.9)$ | 0.004 | 0.474 | -0.089 | 0.097 | 0.001 | 0.579 | -0.112 | 0.115 |
| $\sigma(1.5)$ | - | - | - | - | 0.967 | 0.240 | 0.920 | 1.014 |
| $ICC^4$ | - | - | - | - | 0.481 | 0.092 | 0.463 | 0.499 |
| $\chi^2$ | 201.1 | (df = 86), p-value = 0.000 | | | 71.6 | (df =86), p-value = 0.868 | | |
| $C$ | 38.8 | (df = 8), p-value = 0.000 | | | 13.2 | (df = 8), p-value = 0.102 | | |
| AIC | 398.8 | | | | 357.0 | | | |

1. Standard deviation of estimate; 2. Lower 95% confidence interval; 3. Upper 95% confidence interval; 4. Intraclass correlation.
5. True value of the parameter.

Table 20: Simulation result from data produced by adjacent categories logit random-effects model with ($\sigma = 1.5$) by proportional odds model

| Parameter | No Random Effect | | | | Random Effect | | | |
|---|---|---|---|---|---|---|---|---|
| | Estimate | SD[1] | LCI[2] | UCI[3] | Estimate | SD | LCI | UCI |
| $c_1(-2^5)$ | -1.828 | 0.590 | -1.943 | -1.712 | -2.356 | 0.761 | -2.505 | -2.207 |
| $c_2(-1)$ | -0.770 | 0.569 | -0.882 | -0.659 | -0.977 | 0.741 | -1.122 | -0.832 |
| $m_2(0.1)$ | 0.088 | 0.464 | -0.003 | 0.179 | 0.119 | 0.615 | -0.002 | 0.239 |
| $m_3(-0.2)$ | -0.209 | 0.512 | -0.310 | -0.109 | -0.296 | 0.674 | -0.428 | -0.163 |
| $f_2(0.7)$ | 0.563 | 0.550 | 0.455 | 0.671 | 0.756 | 0.740 | 0.611 | 0.901 |
| $f_3(0.6)$ | 0.576 | 0.527 | 0.472 | 0.679 | 0.768 | 0.692 | 0.632 | 0.903 |
| $f_4(1)$ | 0.777 | 0.504 | 0.678 | 0.876 | 1.049 | 0.658 | 0.920 | 1.178 |
| $b_2(0.6)$ | 0.327 | 0.504 | 0.229 | 0.426 | 0.441 | 0.678 | 0.308 | 0.574 |
| $b_3(0.1)$ | 0.581 | 0.540 | 0.475 | 0.687 | 0.772 | 0.720 | 0.630 | 0.913 |
| $b_4(0.9)$ | -0.029 | 0.582 | -0.143 | 0.085 | -0.043 | 0.780 | -0.195 | 0.110 |
| $\sigma(1.5)$ | - | - | - | - | 1.328 | 0.202 | 1.288 | 1.367 |
| $ICC^4$ | - | - | - | - | 0.630 | 0.072 | 0.616 | 0.644 |
| $\chi^2$ | 247.5 | (df = 86), p-value = 0.000 | | | 64.4 | (df =86), p-value = 0.961 | | |
| $C$ | 49.8 | (df = 8), p-value = 0.000 | | | 13.8 | (df = 8), p-value = 0.093 | | |
| $AIC$ | 449.8 | | | | 375.8 | | | |

1. Standard deviation of estimate; 2. Lower 95% confidence interval; 3. Upper 95% confidence interval; 4. Intraclass correlation.
5. True value of the parameter.

Table 21: Simulation result from data produced by adjacent categories logit random-effects model with ($\sigma = 1.5$) by adjacent categories logit model

| Parameter | No Random Effect | | | | Random Effect | | | |
|---|---|---|---|---|---|---|---|---|
| | Estimate | SD[1] | LCI[2] | UCI[3] | Estimate | SD | LCI | UCI |
| $c_1(-2^5)$ | -0.415 | 0.441 | -0.502 | -0.329 | -1.049 | 0.570 | -1.161 | -0.938 |
| $c_2(-1)$ | -1.297 | 0.398 | -1.375 | -1.219 | -1.354 | 0.525 | -1.457 | -1.251 |
| $m_2(0.1)$ | 0.063 | 0.304 | 0.003 | 0.122 | 0.085 | 0.425 | 0.002 | 0.168 |
| $m_3(-0.2)$ | -0.132 | 0.338 | -0.198 | -0.065 | -0.201 | 0.466 | -0.292 | -0.110 |
| $f_2(0.7)$ | 0.377 | 0.362 | 0.306 | 0.448 | 0.531 | 0.518 | 0.430 | 0.633 |
| $f_3(0.6)$ | 0.384 | 0.357 | 0.313 | 0.454 | 0.538 | 0.493 | 0.442 | 0.635 |
| $f_4(1)$ | 0.515 | 0.343 | 0.447 | 0.582 | 0.733 | 0.464 | 0.642 | 0.824 |
| $b_2(0.6)$ | 0.210 | 0.334 | 0.145 | 0.276 | 0.300 | 0.467 | 0.209 | 0.392 |
| $b_3(0.1)$ | 0.375 | 0.353 | 0.306 | 0.444 | 0.528 | 0.494 | 0.431 | 0.624 |
| $b_4(0.9)$ | -0.018 | 0.386 | -0.093 | 0.058 | -0.037 | 0.544 | -0.144 | 0.069 |
| $\sigma(1.5)$ | - | - | - | - | 0.912 | 0.154 | 0.881 | 0.942 |
| $ICC^4$ | - | - | - | - | 0.448 | 0.083 | 0.432 | 0.465 |
| $\chi^2$ | 246.1 | (df = 86), p-value = 0.000 | | | 67.6 | (df =86), p-value = 0.929 | | |
| $C$ | 51.9 | (df = 8), p-value = 0.000 | | | 13.8 | (df = 8), p-value = 0.088 | | |
| $AIC$ | 447.7 | | | | 374.0 | | | |

1. Standard deviation of estimate; 2. Lower 95% confidence interval; 3. Upper 95% confidence interval; 4. Intraclass correlation.
5. True value of the parameter.

Table 22: Simulation result from data produced by adjacent categories logit random-effects model with ($\sigma = 1.5$) by continuation ratio logit model

| Parameter | No Random Effect | | | | Random Effect | | | |
|---|---|---|---|---|---|---|---|---|
| | Estimate | SD[1] | LCI[2] | UCI[3] | Estimate | SD | LCI | UCI |
| $c_1(-2^5)$ | -1.722 | 0.527 | -1.825 | -1.619 | -2.189 | 0.676 | -2.321 | -2.056 |
| $c_2(-1)$ | -1.417 | 0.499 | -1.515 | -1.320 | -1.458 | 0.648 | -1.585 | -1.331 |
| $m_2(0.1)$ | 0.072 | 0.398 | -0.006 | 0.150 | 0.100 | 0.540 | -0.006 | 0.205 |
| $m_3(-0.2)$ | -0.178 | 0.443 | -0.265 | -0.092 | -0.256 | 0.589 | -0.371 | -0.140 |
| $f_2(0.7)$ | 0.486 | 0.486 | 0.391 | 0.581 | 0.648 | 0.645 | 0.522 | 0.774 |
| $f_3(0.6)$ | 0.493 | 0.453 | 0.404 | 0.582 | 0.659 | 0.600 | 0.541 | 0.776 |
| $f_4(1)$ | 0.669 | 0.436 | 0.584 | 0.755 | 0.904 | 0.573 | 0.792 | 1.016 |
| $b_2(0.6)$ | 0.278 | 0.436 | 0.192 | 0.363 | 0.375 | 0.587 | 0.260 | 0.490 |
| $b_3(0.1)$ | 0.491 | 0.463 | 0.400 | 0.581 | 0.658 | 0.625 | 0.535 | 0.780 |
| $b_4(0.9)$ | -0.029 | 0.510 | -0.129 | 0.071 | -0.046 | 0.681 | -0.179 | 0.088 |
| $\sigma(1.5)$ | - | - | - | - | 1.138 | 0.192 | 1.100 | 1.176 |
| ICC[4] | - | - | - | - | 0.556 | 0.083 | 0.539 | 0.572 |
| $\chi^2$ | 250.8 | (df = 86), p-value = 0.000 | | | 77.5 | (df =86), p-value = 0.732 | | |
| C | 45.7 | (df = 8), p-value = 0.000 | | | 13.8 | (df = 8), p-value = 0.087 | | |
| AIC | 453.9 | | | | 390.1 | | | |

1. Standard deviation of estimate; 2. Lower 95% confidence interval; 3. Upper 95% confidence interval; 4. Intraclass correlation.
5. True value of the parameter.

Table 23: Simulation result from data produced by continuation ratio logit random-effects model with ($\sigma = 1.5$) by proportional odds model

| Parameter | No Random Effect | | | | Random Effect | | | |
|---|---|---|---|---|---|---|---|---|
| | Estimate | SD[1] | LCI[2] | UCI[3] | Estimate | SD | LCI | UCI |
| $c_1(-2^5)$ | -1.416 | 0.464 | -1.507 | -1.325 | -1.735 | 0.563 | -1.846 | -1.625 |
| $c_2(-1)$ | -0.648 | 0.441 | -0.735 | -0.562 | -0.785 | 0.538 | -0.891 | -0.680 |
| $m_2(0.1)$ | 0.085 | 0.399 | 0.007 | 0.163 | 0.109 | 0.496 | 0.012 | 0.206 |
| $m_3(-0.2)$ | -0.135 | 0.397 | -0.213 | -0.057 | -0.181 | 0.485 | -0.276 | -0.086 |
| $f_2(0.7)$ | 0.518 | 0.577 | 0.405 | 0.631 | 0.643 | 0.718 | 0.502 | 0.784 |
| $f_3(0.6)$ | 0.482 | 0.509 | 0.382 | 0.581 | 0.590 | 0.632 | 0.466 | 0.714 |
| $f_4(1)$ | 0.719 | 0.478 | 0.625 | 0.812 | 0.892 | 0.599 | 0.774 | 1.009 |
| $b_2(0.6)$ | 0.392 | 0.451 | 0.304 | 0.481 | 0.496 | 0.572 | 0.384 | 0.608 |
| $b_3(0.1)$ | 0.518 | 0.446 | 0.431 | 0.606 | 0.649 | 0.559 | 0.539 | 0.758 |
| $b_4(0.9)$ | 0.004 | 0.507 | -0.095 | 0.104 | 0.007 | 0.632 | -0.117 | 0.131 |
| $\sigma(1.5)$ | - | - | - | - | 1.130 | 0.202 | 1.091 | 1.170 |
| ICC[4] | - | - | - | - | 0.552 | 0.086 | 0.535 | 0.569 |
| $\chi^2$ | 207.4 | (df = 86), p-value = 0.035 | | | 61.4 | (df =86), p-value = 0.979 | | |
| C | 43.0 | (df = 8), p-value = 0.000 | | | 13.4 | (df = 8), p-value = 0.099 | | |
| AIC | 422.9 | | | | 370.3 | | | |

1. Standard deviation of estimate; 2. Lower 95% confidence interval; 3. Upper 95% confidence interval; 4. Intraclass correlation.
5. True value of the parameter.

**Table 24: Simulation result from data produced by continuation ratio logit random-effects model with ($\sigma = 1.5$) by adjacent categories logit model**

| Parameter | No Random Effect | | | | Random Effect | | | |
|---|---|---|---|---|---|---|---|---|
| | Estimate | SD[1] | LCI[2] | UCI[3] | Estimate | SD | LCI | UCI |
| $c_1(-2^5)$ | 0.194 | 0.347 | 0.126 | 0.262 | -0.176 | 0.412 | -0.257 | -0.095 |
| $c_2(-1)$ | -1.446 | 0.287 | -1.502 | -1.390 | -1.440 | 0.343 | -1.507 | -1.372 |
| $m_2(0.1)$ | 0.050 | 0.242 | 0.002 | 0.097 | 0.068 | 0.311 | 0.007 | 0.129 |
| $m_3(-0.2)$ | -0.082 | 0.238 | -0.128 | -0.035 | -0.112 | 0.299 | -0.170 | -0.053 |
| $f_2(0.7)$ | 0.315 | 0.348 | 0.247 | 0.383 | 0.403 | 0.448 | 0.316 | 0.491 |
| $f_3(0.6)$ | 0.292 | 0.308 | 0.232 | 0.353 | 0.372 | 0.395 | 0.294 | 0.449 |
| $f_4(1)$ | 0.434 | 0.291 | 0.377 | 0.491 | 0.557 | 0.375 | 0.484 | 0.631 |
| $b_2(0.6)$ | 0.236 | 0.274 | 0.182 | 0.289 | 0.306 | 0.356 | 0.237 | 0.376 |
| $b_3(0.1)$ | 0.310 | 0.271 | 0.256 | 0.363 | 0.400 | 0.354 | 0.331 | 0.470 |
| $b_4(0.9)$ | 0.006 | 0.308 | -0.055 | 0.066 | 0.005 | 0.398 | -0.073 | 0.083 |
| $\sigma(1.5)$ | - | - | - | - | 0.699 | 0.130 | 0.674 | 0.725 |
| $ICC^4$ | - | - | - | - | 0.326 | 0.079 | 0.310 | 0.342 |
| $\chi^2$ | 207.1 | (df = 86), p-value = 0.000 | | | 66.1 | (df =86), p-value = 0.945 | | |
| C | 43.4 | (df = 8), p-value = 0.000 | | | 13.3 | (df = 8), p-value = 0.102 | | |
| AIC | 422.5 | | | | 371.2 | | | |

1. Standard deviation of estimate; 2. Lower 95% confidence interval; 3. Upper 95% confidence interval; 4. Intraclass correlation.
5. True value of the parameter.

**Table 25: Simulation result from data produced by continuation ratio logit random-effects model with ($\sigma = 1.5$) by continuation ratio logit model**

| Parameter | No Random Effect | | | | Random Effect | | | |
|---|---|---|---|---|---|---|---|---|
| | Estimate | SD[1] | LCI[2] | UCI[3] | Estimate | SD | LCI | UCI |
| $c_1(-2^5)$ | -1.318 | 0.412 | -1.399 | -1.237 | -1.592 | 0.494 | -1.689 | -1.495 |
| $c_2(-1)$ | -1.609 | 0.390 | -1.685 | -1.533 | -1.611 | 0.461 | -1.702 | -1.521 |
| $m_2(0.1)$ | 0.071 | 0.341 | 0.005 | 0.138 | 0.097 | 0.425 | 0.014 | 0.181 |
| $m_3(-0.2)$ | -0.119 | 0.342 | -0.186 | -0.052 | -0.152 | 0.414 | -0.233 | -0.071 |
| $f_2(0.7)$ | 0.444 | 0.506 | 0.345 | 0.543 | 0.546 | 0.617 | 0.425 | 0.667 |
| $f_3(0.6)$ | 0.419 | 0.441 | 0.332 | 0.505 | 0.505 | 0.542 | 0.398 | 0.611 |
| $f_4(1)$ | 0.620 | 0.419 | 0.538 | 0.703 | 0.758 | 0.514 | 0.658 | 0.859 |
| $b_2(0.6)$ | 0.336 | 0.400 | 0.257 | 0.414 | 0.425 | 0.496 | 0.328 | 0.522 |
| $b_3(0.1)$ | 0.438 | 0.385 | 0.363 | 0.514 | 0.549 | 0.481 | 0.455 | 0.643 |
| $b_4(0.9)$ | 0.008 | 0.448 | -0.080 | 0.096 | 0.008 | 0.551 | -0.100 | 0.116 |
| $\sigma(1.5)$ | - | - | - | - | 0.960 | 0.182 | 0.925 | 0.996 |
| $ICC^4$ | - | - | - | - | 0.472 | 0.091 | 0.454 | 0.490 |
| $\chi^2$ | 209.3 | (df = 86), p-value = 0.029 | | | 72.1 | (df =86), p-value = 0.858 | | |
| C | 40.0 | (df = 8), p-value = 0.000 | | | 13.3 | (df = 8), p-value = 0.102 | | |
| AIC | 425.9 | | | | 380.6 | | | |

1. Standard deviation of estimate; 2. Lower 95% confidence interval; 3. Upper 95% confidence interval; 4. Intraclass correlation.
5. True value of the parameter.